\documentclass[fleqn]{POWFI-2018}
\usepackage{graphicx}
\usepackage{subfigure}
\usepackage{amssymb,amsmath}
\begin{document}

\ensubject{subject}

\ArticleType{Article}
\Year{2019}
\Month{January}
\Vol{60}
\No{1}
\DOI{ }
\ArtNo{000000}
\ReceiveDate{, 2019}
\AcceptDate{, 2019}

\title{Fast jet proper motion discovered in a blazar at $z = 4.72$}

\author[1,2]{Yingkang Zhang}{}
\author[1,3]{Tao An}{antao@shao.ac.cn}
\author[4]{S\'andor Frey}{}

\AuthorMark{Zhang YK}
\AuthorCitation{Zhang, An, Frey}

\address[1]{Shanghai Astronomical Observatory, CAS, Nandan Road 80, Shanghai 200030, China}
\address[2]{University of Chinese Academy of Sciences, 19A Yuquan Road, Shijingshan District, 100049 Beijing, China}
\address[3]{Key Laboratory of Radio Astronomy, Chinese Academy of Sciences, 210008 Nanjing, China}
\address[4]{Konkoly Observatory, Research Centre for Astronomy and Earth Sciences, Konkoly Thege Mikl\'{o}s \'{u}t 15-17, H-1121 Budapest, Hungary}

\abstract{
High-resolution observations of high-redshift ($z>4$) radio quasars offer a unique insight into jet kinematics at early cosmological epochs, as well as constraints on cosmological model parameters. Due to the general weakness of extremely distant objects and the apparently slow structural changes caused by cosmological time dilation, only a couple of high-redshift quasars have been studied with parsec-scale resolutions, and with limited number of observing epochs. Here we report on very long baseline interferometry (VLBI) observations of a high-redshift blazar J1430+4204 ($z=4.72$) in the 8 GHz frequency band at five different epochs spanning 22 years. The source shows a compact core--jet structure with two jet components being identified within 3 milli-arcsecond (mas) scale. The long time span and multiple-epoch data allow for the kinematic studies of the jet components. That results in a jet proper motion of  $\mu {\rm (J1)}$ = 0.017$\pm$0.002~mas\,yr$^{-1}$ and $\mu({\rm J2})$=0.156$\pm$0.015~mas\,yr$^{-1}$, respectively. For the fastest-moving outer jet component J2, the corresponding apparent transverse speed is $19.5 \pm 1.9 \,c$. The inferred bulk jet Lorentz factor $\Gamma = 14.6 \pm 3.8$ and viewing angle $\theta = 2.2^{\circ} \pm 1.6^{\circ}$ indicate highly relativistic beaming. The Lorentz factor and apparent proper motion are the highest measured to date among the $z>4$ jetted radio sources, while the jet kinematics is still consistent with the cosmological interpretation of quasar redshifts.
}


\maketitle


\section{Introduction}

Active galactic nuclei (AGNs) are among the most powerful and energetic objects in the Universe. 
It is believed that every AGN harbours a supermassive black hole (SMBH) in its central region.
Because of their enormous energy production and persistent high luminosity, AGNs are excellent laboratories for studying black hole accretion and galaxy evolution across cosmic times. 
High-redshift quasars (HRQs) that represent powerful AGNs in the early Universe constitute a unique sample for evolutionary studies of AGNs as well as of the cosmic environment \cite{1993Natur.361..134K,1999A&A...342..378G}. The most distant AGN known to date is the quasar J1342+0928 at $z=7.54$, corresponding to only 5\% of the current age of the Universe \cite{2018Natur.553..473B}. The number of quasars known at $z > 5.6$ already exceeds 100 (e.g. \cite{2016ApJ...833..222J,2016ApJS..227...11B}). The discovery of extremely HRQs provided strong constraints on the formation history of the first generation of SMBHs \cite{2006ApJ...650..669V,2018Natur.553..473B}.

Among the HRQs, the radio-loud subsample is worth for exploring their relativistic jets and radio morphologies. Jets from radio-loud HRQs are useful probes of the intergalactic and interstellar environment in the very early stages (e.g. the epoch of reionization). Imaging of HRQ jets requires milli-arcsecond (mas) resolution or even higher. This can be achieved by using very long baseline interferometry (VLBI), which is the highest-resolution imaging technique. The fact that only $\sim$ 10\% of the optically detected quasars are radio-loud, and that radio-loud quasars become too weak to be detected by radio telescopes at very large cosmological distances make high-redshift radio-loud quasars much more rare.  The number of radio-detected AGNs with available spectroscopic redshift at $z>4$ is only about 170 \cite{2017FrASS...4....9P}, and the most distant radio-loud quasar to date is J1429+5447 at $z = 6.21$ \cite{2010AJ....140..546W,2011A&A...531L...5F}.

J1430+4204 was discovered as a radio-loud quasar at $z=4.72$ \cite{1998MNRAS.294L...7H}.
Further X-ray and radio observations confirmed its blazar nature with broad-band variability and flat radio spectrum \cite{1998MNRAS.295L..25F,1999MNRAS.308L...6F,2006MNRAS.368..844W}. A distant ($\sim 3.6^{\prime\prime}$) jet component was detected in the northwestern direction of the nucleus from {\em Chandra} X-ray and Very Large Array (VLA) radio observations, making J1430+4204 the most distant quasar being detected with kiloparsec (kpc) scale radio/X-ray jet known so far \cite{2008ASPC..386..462C,2012ApJ...756L..20C}.
The mas-scale radio structure of J1430+4204 has been studied with VLBI since 1996 \cite{1999A&A...344...51P}. With further more sensitive and higher-resolution VLBI observations,  a compact core--jet structure was revealed, with a weak mas-scale jet ejecting towards the west-southwest seen at 2.3, 8.4 and 5~GHz frequencies (e.g. \cite{2007ApJ...658..203H,2012A&A...544A..34P,2012ApJ...756L..20C}). 
Veres et al. \cite{2010A&A...521A...6V} conducted a two-epoch 15-GHz VLBI study of J1430+4204 before and after a major radio flare from 2005 to 2006. They did not find any evidence of newly ejected jet components. The jet emission in their 15-GHz VLBI images appears diffuse, hampering a direct measurement of jet proper motion. 

In this paper, we present a multi-epoch VLBI analysis of J1430+4204. Thanks to the long time span and multiple epochs, we are able to study the jet kinematics of J1430+4204. The five-epoch data were obtained in the 8-GHz frequency band, where the jet components are bright and compact enough for model-fitting. 
In Sect.~\ref{sec2}, we provide the basic information about our VLBI data and describe their analysis. Section~\ref{sec3} presents the high-resolution images, the results of brightness distribution modeling, and the calculation of jet parameters. In Sect.~\ref{sec4}, We discuss the properties of J1430+4204 and put them into context with other extremely distant quasars studied with VLBI. 

Throughout this paper, a flat $\Lambda$CDM cosmological model was adopted, with the parameters of H$_{0}=$ 70 km s$^{-1}$ Mpc$^{-1}$, $\Omega_{\rm m}=$ 0.27, and $\Omega_{\Lambda}=$ 0.73.
At the redshift of J1430+4204, 1~mas angular size corresponds to 6.68~pc projected linear size and 1~mas\,yr$^{-1}$ proper motion to $124.7 c$ apparent transverse speed ($c$ denotes the speed of light) \cite{2006PASP..118.1711W}.

\section{Methods}
\label{sec2}

Data used for the jet kinematic study were obtained from VLBI observations in a total of five epochs from 1996 June 8 to 2018 May 1 (more details are referred to \ref{app:1}). All these observations were made in geodetic/astrometric mode, in which a target source is observed during a few scans, usually for 5 min in each scan. The details of the observation logs are presented in Table \ref{tab1}. Except for the fourth epoch observed only at 8.4 GHz, the other observations were carried out at dual 2/8 GHz frequency bands. The image noise depends on a combination of factors such as on-source integration time, number of telescopes and their diameters, data rate, (u,v) coverage, quality of calibration, etc. The final root-mean-square (rms) noise in the naturally-weighted {\sc clean} images ranges from 0.25 to 1.27 mJy beam$^{-1}$. The visibility data were calibrated using the US National Radio Astronomy  Observatory  (NRAO)  Astronomical  Image  Processing  System  ({\sc AIPS})  software package \cite{2003ASSL..285..109G}. The calibrated VLBI data were imported into the {\sc difmap} software package \cite{1997ASPC..125...77S} to carry out self-calibration, imaging and model fitting (\ref{app:2}).

\section{Results}
\label{sec3}

\subsection{Parsec-scale radio jet}

\begin{figure}
  \centering
  \includegraphics[width=8cm, angle=0]{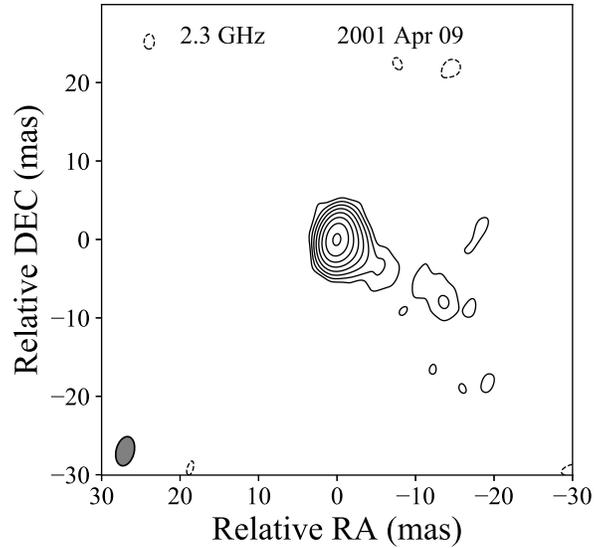}
  \caption{Our best-quality VLBI image of J1430+4204 with natural weighting at 2.3~GHz (epoch 2001 April 9). The lowest contours are at $\pm 3\sigma$ image noise level, the positive contour levels increase by a factor of 2. The peak brightness is 114.9~mJy\,beam$^{-1}$. More image parameters can be found in Table~\ref{tab1}.} 
  \label{Fig1}
\end{figure}

Because of the limited sensitivity of geodetic-style `snapshot' VLBI experiments with short on-source integration times and poor (u,v) coverage (see Table~\ref{tab1}), not all epochs are sufficient to produce a reasonably detailed image of the core--jet structure. At 2.3~GHz band, we show only the best-quality image made from the epoch 2001 data (Fig.~\ref{Fig1}) to indicate that the jet emission continues expanding out to an extent $\sim 20$~mas (or a projected distance of $\sim$130 pc) from the core (the brightest component at the image center) in the southwest direction. Our 2.3-GHz VLBI image (Fig.~\ref{Fig1}) is qualitatively consistent with that obtained in 1998 \cite{2012ApJ...756L..20C}. Both images reveal a compact core--jet structure within 10~mas and weaker knotty components beyond 10~mas. 

\begin{figure}
   \centering
  \includegraphics[width=5cm, angle=0]{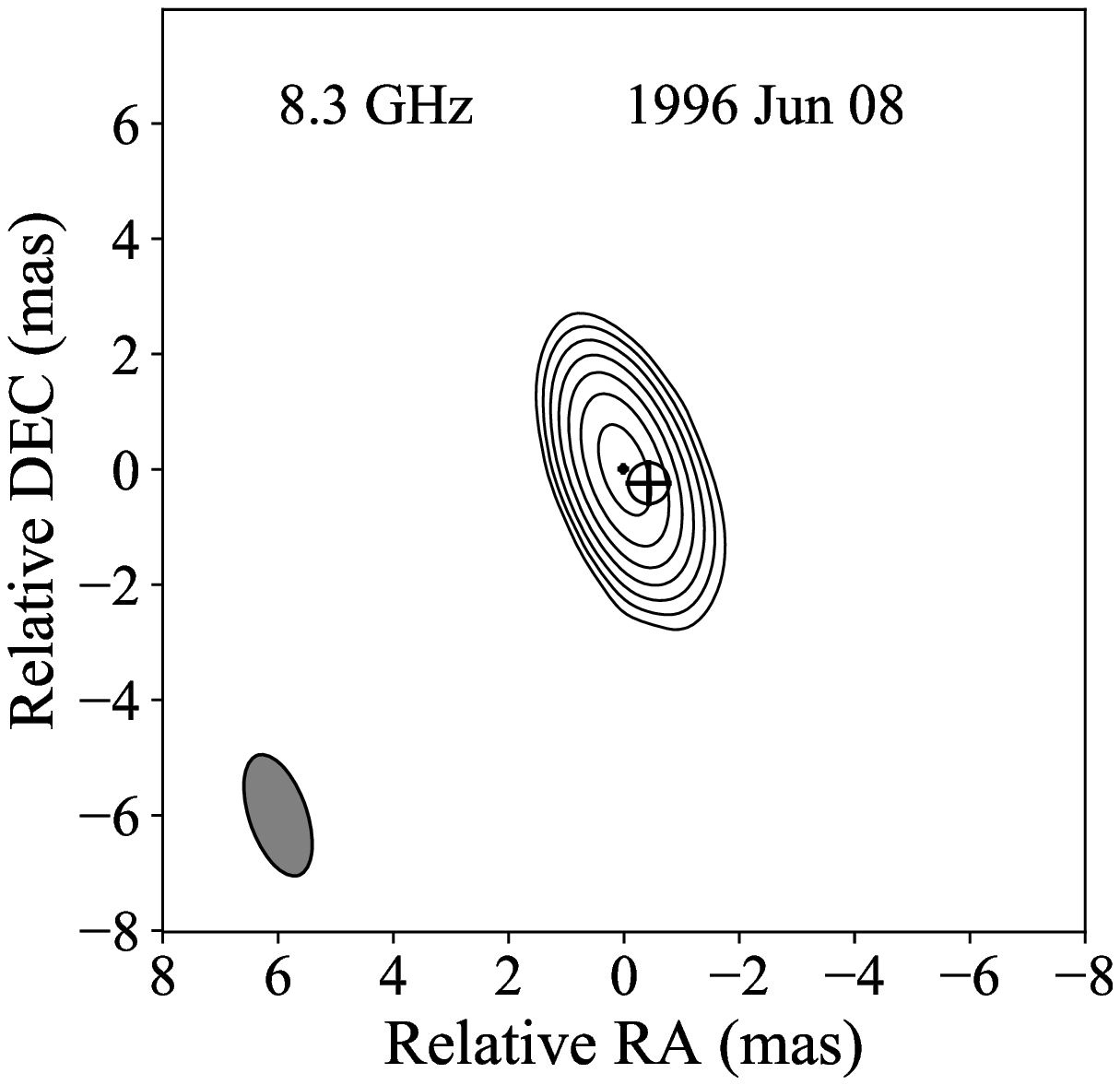}
  \includegraphics[width=5cm, angle=0]{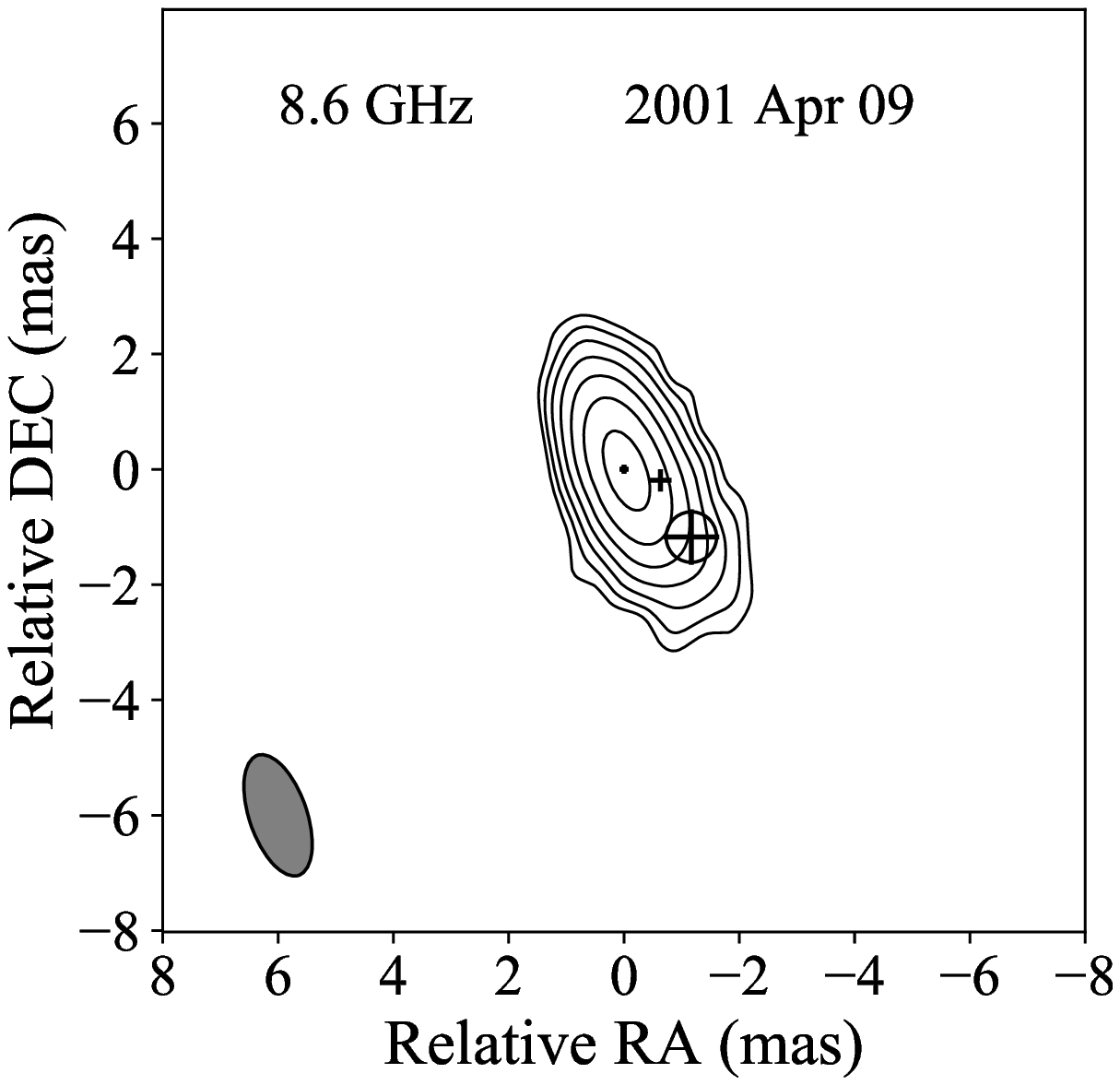}
  \includegraphics[width=5cm, angle=0]{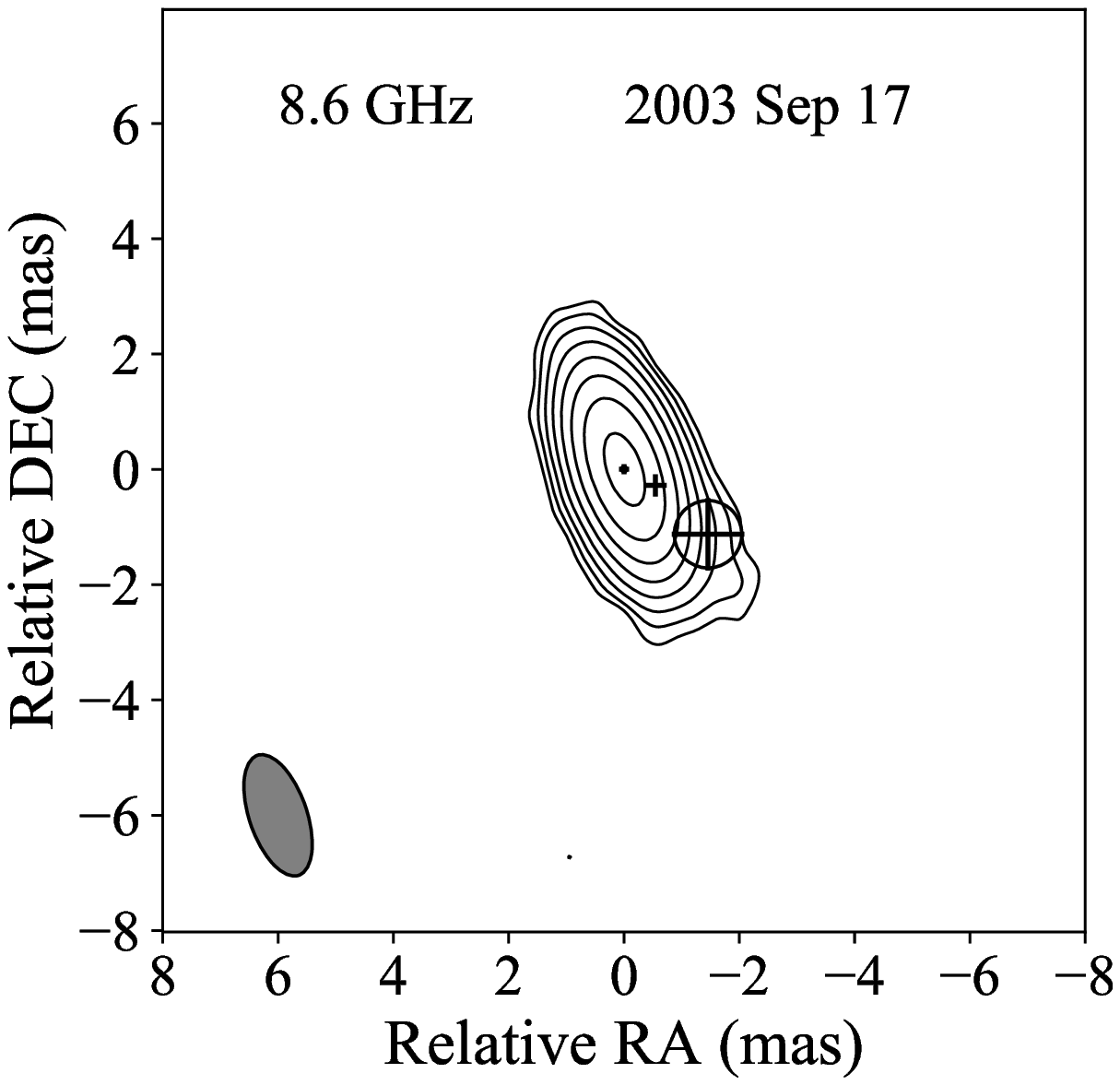}
  \includegraphics[width=5cm, angle=0]{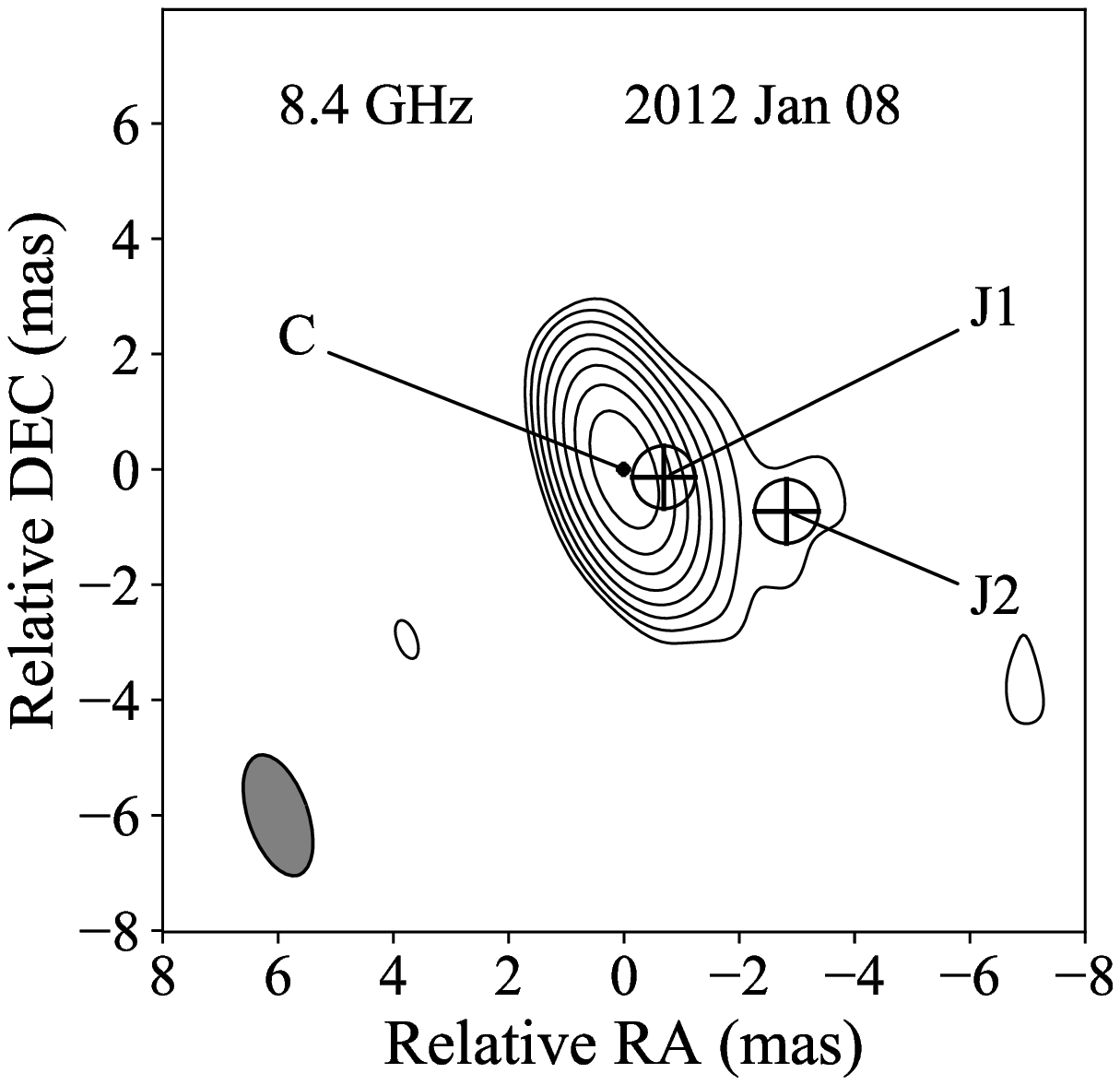}
  \includegraphics[width=5cm, angle=0]{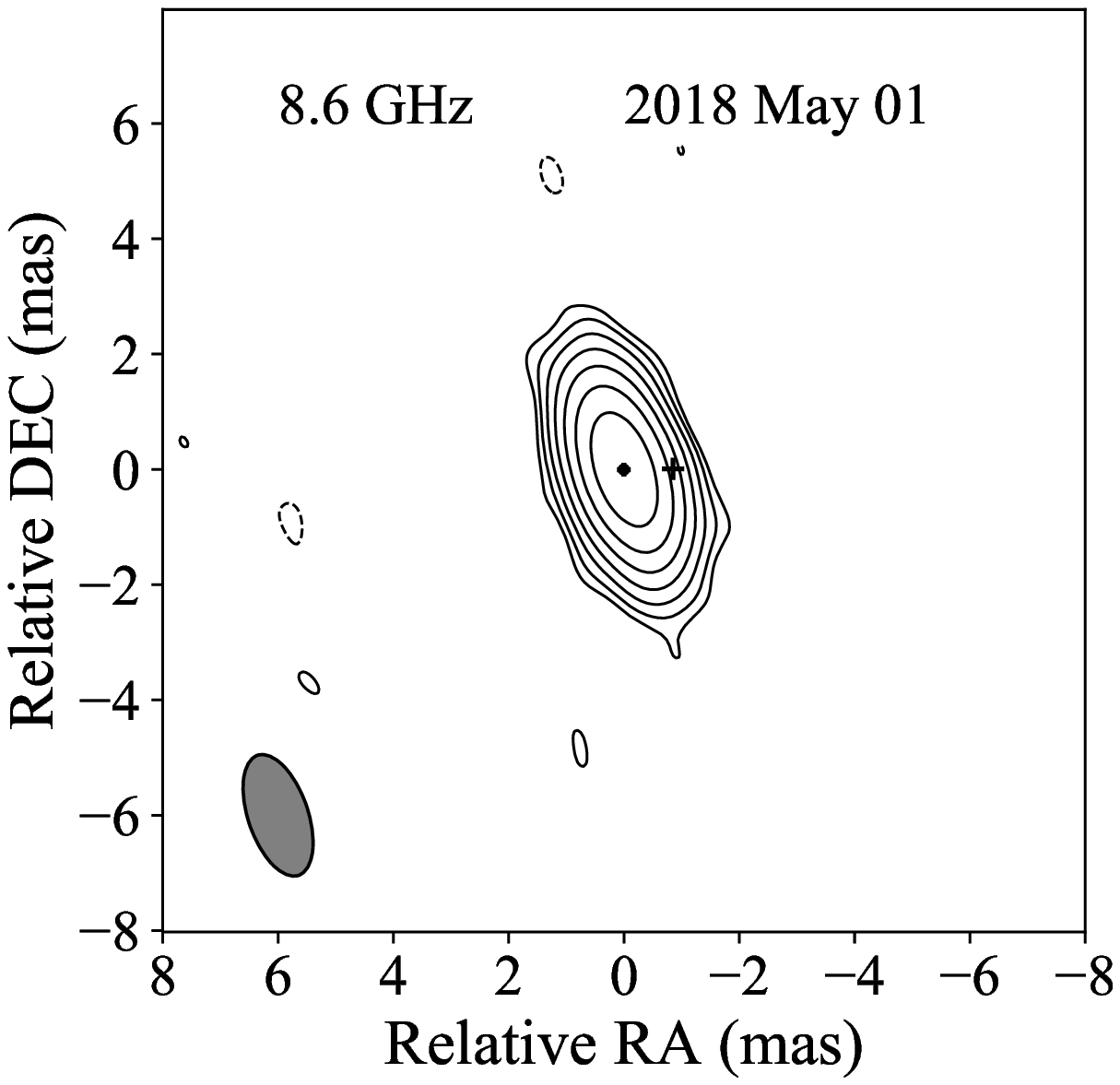}
   \caption{Naturally weighted VLBI images of J1430+4204 at 8~GHz at the five  epochs. Basic information of the maps can be found in Table~\ref{tab1}. For a consistent presentation of the images, we restored all of them with the largest beam corresponding to the 2012 epoch. The restoring beam (FWHM) is shown at the bottom-left corner of each image. The lowest contours are at $\pm 3\sigma$ level of individual images and the positive contour levels increase by a factor of 2. The peak intensities are 181.4, 69.7, 156.6, 182.3, and 168.3 mJy\,beam$^{-1}$, respectively. In each image, the circles with a cross inside represent the locations and FWHM diameter of the fitted circular Gaussian model component; the cross symbols represent the positions of point models.}
   \label{Fig2}
\end{figure}

Figure \ref{Fig2} shows the 8-GHz images at five available epochs. 
The 8-GHz imaging observations resolve out the diffuse emission structure at $>$10-mas scale and show jet components only within about 3~mas from the core (Fig.~\ref{Fig2}). The core showed a significant flux density variability from 182.3 mJy at epoch 1996 to 65.7 mJy at epoch 2001, then to 152.8 mJy at epoch 2003.  The jet components are located to the west and southwest of the core, roughly in alignment with the 2.3-GHz jet structure (Fig. \ref{Fig1}). The VLA image at 1.4 GHz made by Cheung et al. \cite{2012ApJ...756L..20C} implies an extended jet/lobe at $3.6^{\prime\prime}$ in the northwest. 
A large difference of position angles ($\sim 50^\circ$) between the mas- and arcsec-scale jet indicates a possible large jet bending. Alternatively, the northwest lobe could be a relic from the past cycle of radio activity. To identify the connection between the VLBI and VLA jets, intermediate-resolution ($\sim 100$ mas) radio interferometric imaging would be required.  

\subsection{Jet proper motion}

The typical one-sided apparent jet morphology of blazars provide a unique opportunity for their kinematic study, where the jet component proper motions, the viewing angles and the Lorentz factor of the relativistic jet can be constrained using high-resolution VLBI observations.
Due to the cosmological time dilation effect, structural changes in the jet are seen $(1+z)$ times slower in the observer's frame than in the rest frame of the high redshift quasars \cite{2015MNRAS.446.2921F}. Thus with an assumption of a continuous jet flow, we are able to link the jet components with each other at five different epochs over an observed period spanning $\sim 22$~yr that actually corresponds to just about 4~yr in the rest frame of J1430+4204 at $z=4.72$.

The 2.3 GHz observing frequency does not provide enough angular resolution to distinguish between the innermost jet components. Moreover, the outermost component seen at 2.3~GHz is too faint and diffuse, preventing us from making reliable proper motion estimates. Therefore the jet kinematic analysis is based on the 8-GHz observations. 
With model-fitting to our VLBI data, we were able to identify two jet components, named J1 and J2, across the different epochs. Component J1 is located well within 1~mas from the core and was detected at all 5 epochs. Component J2 shows an apparent separation of $\sim2-3$~mas from the core and was detected in the middle 3 epochs.
The first and fifth epochs have relatively lower imaging sensitivity (see column 9 in Table B1) due to the smaller number of telescopes (in epoch 5) and short integration time (in epoch 1). These lead to a non-detection of the weaker jet component J2.
The parameters of individual components derived from model-fitting are listed in Table~\ref{tab2}. The locations and sizes of the fitted model components are also indicated in Fig.~\ref{Fig2}.

From the parameters in Table~\ref{tab2}, we found that both jet components are moving outwards from the core. 
The trajectory of J2 appears slightly curving towards the north (i.e. counter-clockwise direction in Fig.~\ref{Fig2}), with the jet position angle gradually changing from $-135.2^\circ$ at epoch 2001 to $-104.4^\circ$ at epoch 2012. 
The component J1 also shows a gradual change of position angle. 
Small changes in the direction of a relativistic jet flow beaming toward the observer could be amplified by projection effect.  
For example, apparent jet curvature could be attributed to a low-pitch helical trajectory projected on the sky \cite{1993ApJ...411...89C,1993AAS...18310705C}, which is commonly seen in many blazars \cite{2010ApJ...710..764K,2011A&A...529A.113Z,2013AJ....146..120L}. The northward rotation of the jet in our VLBI images might be related to the larger, kpc-scale radio structure of J1430+4204, where a distant jet component at $3.6^{\prime\prime}$ from the core is seen in the VLA image \cite{2012ApJ...756L..20C}. An alternative possibility of a jet deflection off the dense interstellar medium cannot be ruled out \cite{1997A&A...318...11G,2000Sci...289.2317G,2017MNRAS.468.2699L}. Further polarisation-sensitive VLBI observations and long-term flux density monitoring to reveal periodicity could address the problem of the helical jet.  

\begin{figure} 
	\centering
	\includegraphics[width=8.0cm, angle=0]{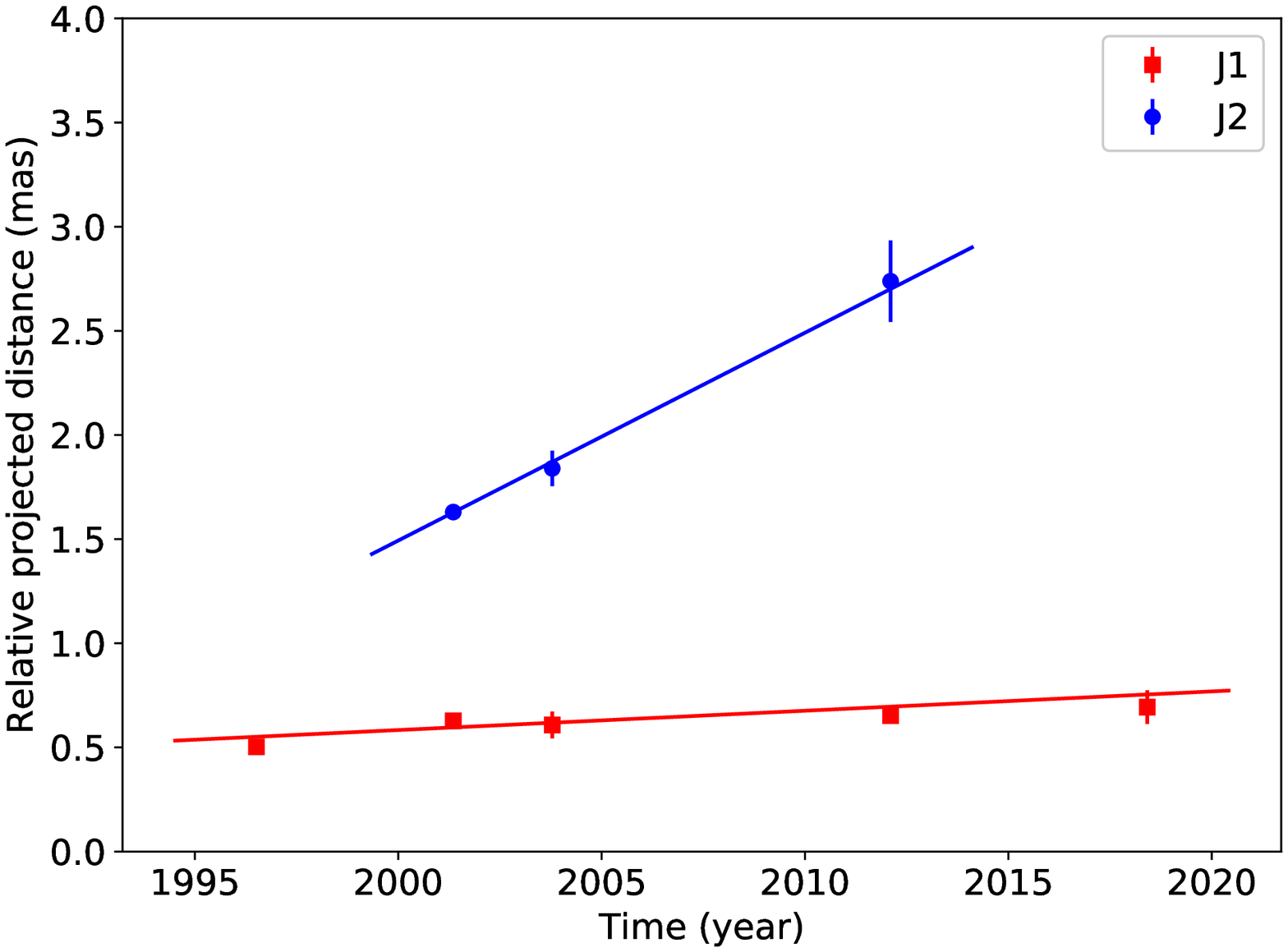}
	\includegraphics[width=8.0cm, angle=0]{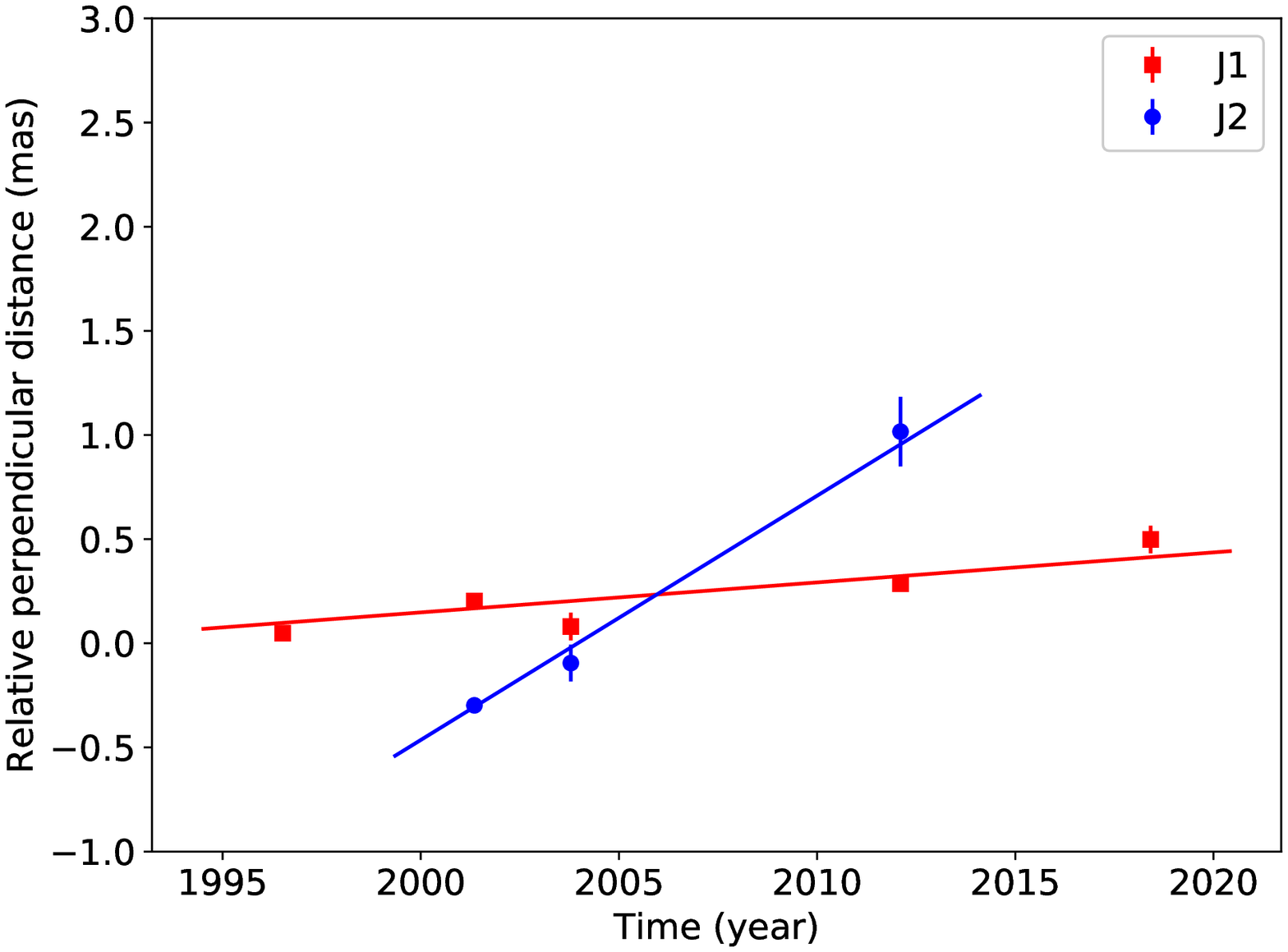}
	\caption{Component positions with respect to the core as a function of time and the fitted proper motions for J1 and J2. The left panel shows the projected motions along the 2.3-GHz jet direction. The right panel shows the motions perpendicular to the 2.3-GHz jet direction from southeast to northwest. The fitted parameter values can be found in Table~\ref{tabpm}.} 
	\label{Fig3}
\end{figure}

To describe the apparent motions of J1 and J2 relative to the core, we estimated their linear proper motions along the 2.3- jet direction and perpendicular directions, respectively, using least-squares fitting. The component position angles from the best-resolved 2.3-GHz image (Fig.~\ref{Fig1}) characterize a relativistic jet moving to southwest, where the mean position angle is $-124.8^\circ$. This value is adopted to define the pc-scale jet direction here. Figure~\ref{Fig3} shows the fitted jet proper motions with respect to the core along and perpendicular to this direction. The values are $\mu_{||} = $ 0.100 $\pm$ 0.017 mas yr$^{-1}$ and $\mu_{\perp} = $0.120 $\pm$ 0.014 mas yr$^{-1}$ for J2, $\mu_{||} = $ 0.009 $\pm$ 0.002 mas yr$^{-1}$ and $\mu_{\perp} = $0.015 $\pm$ 0.002 mas yr$^{-1}$ for J1, respectively. Veres et al. \cite{2010A&A...521A...6V}  studied this source based on two-epoch 15-GHz VLBA observations. They estimated the jet parameters but did not detect a moving component due to their short observation period. Another reason is that the jet emission at 15~GHz (corresponding to 86~GHz in the source rest frame) is optically thin and becomes intrinsically weak.

Veres et al. \cite{2010A&A...521A...6V} estimated the Doppler factor using three independent approaches: from radio flux density variability, from radio and X-ray data and the inverse Compton process, and from the core brightness temperature measured with VLBI at 15~GHz. A moderate bulk Lorentz factor of 4.6--6.8 was derived from their VLBI study.
We also applied the core brightness temperature to estimate the jet beaming parameters, while obtained a higher Lorentz factor $\Gamma = 14.6\pm3.8$ and a viewing angle of $2.2^\circ\pm1.6^\circ$ (\ref{app:3}).
Our calculated jet parameters are in agreement with the blazar nature of J1430+4204 whose relativistic jet is inclined within a very small angle to the line of sight.
The 15 GHz (correponding to a source-rest-frame frequency of about 86 GHz)  data of Veres et al. \cite{2010A&A...521A...6V} probe an inner jet section, while our 8 GHz data trace a relatively outer region. The difference in Lorentz factors derived from Veres et al. and ours may imply an increase of the jet flow Lorentz factor from a distance $\sim$100 pc (deprojected distance of J1) to $\sim$500 pc (deprojected distance of J2). 

\section{Discussion and conclusion}
\label{sec4}
According to the unified scheme of radio-loud (jetted) AGNs, the appearance and luminosity of the objects depend on their jet orientation with respect to the viewing direction \cite{1995PASP..107..803U}. Blazars, whose relativistic jets point nearly towards us, are intriguing sources to probe the early Universe. Their jet emission is strongly enhanced due to the Doppler beaming effect which makes them more easily detectable compared to the coetaneous quasars with unbeamed jets, especially at high redshifts (e.g. \cite{2011MNRAS.416..216V}). Moreover, the powerful aligned jets are not affected by the obscuring torus that surrounds the central black hole and have more chances to break through the surrounding material \cite{2016MNRAS.461L..21G}, making the high-$z$ sources prominent in multi-band studies \cite{2011ApJ...736...57Z,2019MNRAS.489.2732I}. 

\begin{table}
	\begin{minipage}[]{100mm}
		\caption[]{Jet kinematic parameters of $z>4$ quasars derived from previous VLBI studies.}\label{tab4}\end{minipage}
	\setlength{\tabcolsep}{2.5pt}
	\centering
	\begin{tabular}{ccccccc}
		\hline
		Name & redshift & $\beta$ & $\delta$ & $\Gamma$ & $\theta$ & Ref. \\
		(1)  &   (2)    &   (3)   &   (4)    &   (5)    &   (6)    & (7) \\
		\hline 
		J0906$+$6930& 5.47& 2.5$\pm$0.8  & 6.0$\pm$0.8 & 3.6$\pm$0.5 &6.8$\pm$2.2$^{\circ}$  &1\\
		J1026$+$2542& 5.27& 12.5 $\pm$ 3.9  & $\gtrsim$ 5 &$\gtrsim$ 12.5  &4.6$^{\circ}$& 2\\
		J1430$+$4304& 4.72& 2--4.2  & 8.6--12 & 4.6--6.8 & 3$^{\circ}$ $^*$ & 3 \\
		J2134$-$0419& 4.33& 4.1 $\pm$ 2.7  & 3--8 &2--7  & 3--20$^{\circ}$ & 4 \\
		\hline
		J1430$+$4304& 4.72& 19.5$\pm$1.9  & 22.2$\pm$10.6 & 14.6$\pm$3.8 & 2.2$\pm$1.6$^{\circ}$ & 5\\ 
		\hline
	\end{tabular}
	\\
	{Col.~3 --Fastest apparent transverse speed in unit of $c$; Col.~4 -- Doppler factor; Col.~5 -- bulk Lorentz factor; Col.~6 -- jet viewing angle with respect to our line of sight; Col.~7 -- reference: 
	1. \cite{An2019NC}; 
	2. \cite{2015MNRAS.446.2921F}; 
	3. \cite{2010A&A...521A...6V}; 
	4. \cite{2018MNRAS.477.1065P}; 
	5. This paper
\newline $^*$ this value is not estimated but fixed to an assumed value of $3^\circ$.}
\end{table}

Apparent proper motion estimates of AGN jets provide a unique way of understanding their relativistic motions. This requires multi-epoch VLBI observations at a given frequency. The cosmological time dilation effect makes it more difficult to see apparent changes in high-redshift radio jets than in low-redshift ones: any change is observed $(1+z)$ times slower compared to the rest frame of the quasar. Therefore determination of jet proper motion in HRQs has to possess long enough gaps from epoch to epoch. Among the known radio quasars at $z > 4.5$, J1430+4204 belongs to the most luminous ones \cite{2016MNRAS.463.3260C}, owing to its Doppler-boosted jet emission. 
Until now, direct estimates of AGN jet proper motion at $z>4$ are rare, available only for 3 objects: J0906+6930 ($z=5.47$) at 15~GHz (An et al. 2019, in prep.), J1026+2542 ($z=5.26$) at 5~GHz \cite{2015MNRAS.446.2921F}, and J2134$-$0419 ($z=4.33$) at 5~GHz \cite{2018MNRAS.477.1065P}.
Their derived jet parameters are collected in Table~\ref{tab4}, along with our new results obtained for J1430+4204. Compared with previous proper motion estimates for other HRQs, our results are derived from more epochs over a longer time range, making the structural changes more evident and the estimates more reliable. Our estimated vector proper motion of $\sim 0.156$~mas\,yr$^{-1}$ ($\sim 19.5\,c$) for the component J2 stands as the highest among all the measured values for other $z > 4$ quasars. In addition,  a $\gamma$-ray flare was also reported to originate from this source \cite{2018ApJ...865L..17L}, indicating that J1430+4204 is an exceptionally powerful blazar in the early Universe and deserves comprehensive multi-band studies. 

HRQ jets open a window to explore the earliest activity of the accreting SMBHs \cite{2012ApJ...756L..20C}. In addition, the jet parameters (e.g., bulk Lorentz factor, viewing angle) yield useful constraints on the X-ray jet emission mechanism of high-redshift quasars \cite{2003ApJ...598L..15S}.
The resulting $\mu$ and $\Gamma$ of J1430+4204 are at the higher end compared to the known high-$z$ jets, but sit in the middle of the large low-redshift blazar samples \cite{2004ApJ...609..539K,2008A&A...484..119B,2009AJ....138.1874L}. 
The reason why (the most powerful) high-redshift quasar jets do not have extremely high Lorentz factors (as some of the low-redshift blazars) is still an open question and needs more investigations on larger samples. 

As the kinematic analysis of VLBI jets in the large sample of the 15-GHz MOJAVE (Monitoring of Jets in Active Galactic Nuclei with VLBA Experiments) survey indicates, accelerations as well as non-ballistic component motions are common in powerful AGN jets \cite{2015ApJ...798..134H,2016AJ....152...12L}.
An increase in the Lorentz factor dominates the parallel accelerations which usually take place from the origin until a distance of $10^2$ pc away from the core \cite{2015ApJ...798..134H}, as may also account for the observed increasing jet speed from J1 to J2 in J1430+4204. 
Considering that the jet knot may trace a curved path (see Sect. 3.1), an alternative possibility of the apparent jet acceleration might be due to the jet bending; for a highly beamed jet pointing toward us, any small change in the jet inclination angle may cause a large change of the apparent jet speed, even the intrinsic jet bulk speed remains constant.

Our current study helps accumulating kinematic data on high-redshift radio jets. Eventually, with a sufficiently large sample, the possible evolution of jetted radio sources on cosmological time scales, as well as cosmological model parameters could be constrained with such data \cite{1988ApJ...329....1C,2004ApJ...609..539K}.
The number of blazars at high redshifts can also help constrain the space density of radio-loud high-$z$ AGNs. That yields the information on the space distribution of radio sources and cosmological evolution of the number density. If one assumes that AGN jets point in any direction with equal probability, and define blazars with a viewing angle $\theta < 1/\Gamma$, where $\Gamma$ is the bulk Lorentz factor of the emitting jet plasma, the total number of radio-loud AGNs could be obtained as $N_{\rm total}$ = $N_{\rm blazars} \times 2\Gamma^{2}$ \cite{2014MNRAS.440L.111G,2019MNRAS.484..204C}. Taking the highest Lorentz factor of J1430+4204 into account, that implies a total number of $\sim 450$ radio-loud AGNs at the redshift of 4.72. This number is much higher than that currently detected. High-sensitivity all-sky surveys using the next generation radio telescopes, such as the Square Kilometre Array (SKA) and the next-generation VLA (ngVLA), may boost the detection of radio-emitting HRQs  \cite{2015aska.confE.143P,2018ASPC..517..421M}, allowing for more sophisticated study of the cosmological evolution of radio sources.

\section*{Acknowledgements}
This work is supported by SKA pre-research funding granted by the Ministry of Science and Technology of China (MOST, No 2016YFE0100300), the Chinese Academy of Sciences (CAS), and the Hungarian National Research, Development and Innovation Office (grant 2018-2.1.14-T\'ET-CN-2018-00001).
The authors acknowledge the use of Astrogeo Center database of brightness distributions, correlated flux densities, and images of compact radio sources produced with VLBI.
YZ thanks Shu Fengchun, Alexey Melnikov, Jamie McCallum, Bo Xia for providing the Asia-Oceania VLBI (AOV) data and auxiliary telescope system temperature files.

\section*{Disclosure of potential conflict of interest}
The authors declare no potential conflict of interest.

\section*{Data availability}
All data used in this study are public and can be accessed through the different data archives of the various instruments. NRAO VLBA archive: \url{https://archive.nrao.edu/archive/advquery.jsp}. The authors can provide data supporting this study upon request.

\section*{Code availability}
Upon reasonable request the authors will provide all code supporting this study. Astronomical Image Processing System ({\sc AIPS}) software can be found at \url{http://www.aips.nrao.edu/index.shtml}. {\sc Difmap} software can be found at  \url{ftp://ftp.astro.caltech.edu/pub/difmap/}.

\section*{References}


\appendix

\section{VLBI observations} \label{app:1}

A total of five epochs of VLBI data are analysed in this study. The observations were conducted in several different geodetic/astrometric VLBI projects, spanning nearly 22~yr. Table~\ref{tab1} summarizes the basic information of each observation. Calibrated interferometric visibility data for the first four epochs are taken from the Astrogeo database\footnote{Maintained by L. Petrov, \url{http://astrogeo.org/vlbi_images/}}, which is the largest collection of publicly available VLBI imaging data, mostly from the Very Long Baseline Array (VLBA) Calibrator Surveys (e.g. \cite{2016AJ....151..154G}) and a series of geodetic and astrometric projects (e.g. the Research \& Development VLBA project, \cite{2009JGeod..83..859P}). In the below sections, we only introduced the observation and data calibration of the last epoch.

Data for the epoch 2018 May 1st were observed in the framework of the newly-established Asia--Oceania VLBI project (AOV, \cite{2019ivs..conf..131M}) which observes a sample of bright extragalactic radio sources with not only geodetic and astrometric goals, but also for other scientific purposes such as characterising the Sun's corona and testing general relativity (e.g. \cite{2018A&A...618A...8T,2019arXiv191010529A}). The high-redshift blazar J1430+4204 was observed in the session aov22 (Table~\ref{tab1}), with a global network consisting of ten radio telescopes providing baseline lengths up to $\approx 10\,000$~km. The participating antennas were Badary (32~m diameter), Svetloe (32~m), Zelenchukskaya (32~m) from the Russian ``QUASAR'' VLBI network \cite{SHUYGINA2019150}; Kunming (40~m), Sheshan (25~m) from China VLBI network \cite{An2018}; Hobart (26~m), Katherine (12~m), Yarragadee (12~m) from Australian VLBI network \cite{2017JGeod..91..803P}; Warkworth (12~m) from New Zeland, and Ishioka (13~m) from Japan.
For our target J1430+4204, six scans of 6--8~min were scheduled in this session, spreading over 12~h, and providing a good (u,v) coverage for image reconstruction. A total of 16 intermediate frequency channels (IFs) were used in right-hand circular polarization, with 6 IFs at S band (2.3~GHz) and 10 IFs at X band (8.6~GHz), respectively. Each IF was divided into 32 spectral channels of 500~kHz width, resulting in a total bandwidth of 256~MHz. After observation, the data from each station were transported to the Shanghai Astronomical Observatory (China) for correlation, using the DiFX correlator system \cite{2011PASP..123..275D}.

\section{Data processing} \label{app:2}

The phases and amplitudes of the correlated visibility data were calibrated using the US National Radio Astronomy Observatory (NRAO) Astronomical Image Processing System ({\sc AIPS}) software \cite{2003ASSL..285..109G}, following a standard procedure (e.g. \cite{1995ASPC...82..227D}). 
The data were downloaded on the China SKA Regional Centre prototype \cite{2019NatAs...3.1030A} in which the calibration and imaging were carried out. 
After loading and inspecting the correlated data, we separated the S- and X-band data into individual files for further calibration. The amplitudes were calibrated using the system temperatures measured during the observation and known antenna gain curves from each station.

The manual phase calibration and global fringe-fitting were done using the {\sc AIPS} task {\sc FRING}, with the aim of correcting instrumental delays as well as the errors on delays, delay rates and phases. We picked the bright radio source OJ\,287 also observed in the same aov22 experiment, to conduct the manual phase calibration. The fringe-fitted data of OJ\,287 were exported to the Caltech {\sc Difmap} software \cite{1997ASPC..125...77S} for imaging. After a few iterations of hybrid mapping cycles (e.g. \cite{1980Natur.285..137R,1986AJ.....92..213L}), the gain correction factors for each antenna could be determined through the {\sc gscale} command. The antenna-based gain correction factors derived from the bright calibrator OJ\,287 were then applied to the data in AIPS, which refined the amplitude scales that were mainly based on the system temperatures measured at individual antenna sites. 

Since the target source J1430+4204 is bright enough, we performed direct fringe-fitting in {\sc AIPS}. Quite a few antennas in our AOV session were small in diameter and therefore less sensitive than the antennas used at previous epochs. This resulted in weak fringes for the target source and caused larger rms image noise, as indicated in Table~\ref{tab1}. Finally, the calibrated visibility data of J1430+4204 were averaged in frequency within each IF and in time over 2~s, and loaded into {\sc Difmap} for imaging and model-fitting.

\begin{table}
\begin{minipage}[]{100mm}
\caption[]{VLBI observation log\label{tab1}}\end{minipage}
\setlength{\tabcolsep}{1pt}
\small
\centering
 \begin{tabular}{cccccccrccc}
 	\hline
 	\noalign{\smallskip}
Project code &   Date   & Frequency &  Time  &  BW   & $B_{\rm maj}$ & $B_{\rm min}$ &   $B_{\rm PA}$   &    $\sigma$     &           Array            & Ref. \\
 	         & YYYYMMDD &   (GHz)   & (min) & (MHz) &   (mas)   &   (mas)   & ($^{\circ}$) & (mJy\,beam$^{-1}$) &                            &  \\ 
 	     (1) &(2)&(3)&(4)&(5)&(6)&(7)&(8)&(9)&(10)&(11) \\    \hline
 	\noalign{\smallskip}
	bb023   & 19960608 &  8.3  &   5    &  32   &   1.9    &  0.9  &   $-$8.5           &   0.66    &            VLBA            &  1   \\
	... & ... & 2.3 & ... & ... & 6.9 & 3.2 &$-$8.8 & 0.90 & ... & ... \\
       rdv27     & 20010409 &  8.6  &   40   &  32   &   1.0    &   0.6    &  $-$15.7           &    0.27    &    VLBA$+$GcKkMcNyTsWz     &  2   \\
     ... & ... & 2.3  &...&...&3.8&2.3&$-$14.1&0.27 & ... & ... \\
       rdv41& 20030917 &  8.6  &   40   &  32   &   0.8    &   0.6   &      2.1         &    0.31    & VLBA (except BR)$+$GcKkNyTsWfWz &  2   \\
       ...& ...& 2.3&...&...&3.0&2.3&$-$8.0&0.39&VLBA (except BR)$+$GcKkNyTsWfWzOn & ... \\
 bc196zo    & 20120108 &    8.4    &  5.5   &  128  &   2.2    &   1.0    &      18.9        &    0.25    &       VLBA (except BR MK)        &  1   \\
 aov22 & 20180501 & 8.6	 & 50  & 160 & 1.1& 0.6 & 31.1 & 0.48 & BdIsKmShSvZc & 3 \\
 ...&...&2.3&...&96&3.7&2.6&32.7&1.27&BdIsKmShSvZcKe&...\\
 	   \noalign{\smallskip}\hline     &
 \end{tabular}
{Col.~4 -- integration time; Col.~5 -- total observing bandwidth; Cols.~6-7 -- major and minor axis sizes of the restoring beam (FWHM); Col.~8 -- position angle of the major axis of the restoring beam, measured from north to east; Col.~9 -- root-mean-square (rms) noise of the residual map;  Col.~10 -- antennas participated in the VLBI observation: the Very Long Baseline Array (VLBA) of the USA generally contains the following ten 25-m telescopes: Brewster (BR), Fort Davis (FD), Hancock (HN), Kitt Peak (KP), Los Alamos (LA), Mauna Kea (MK), North Liberty (NL), Owens Valley (OV), Pie Town (PT), and Saint Croix (SC). The extra antennas are: Gilmore Creek in the USA (Gc); Kokee Park in the USA (Kk); Medicina in Italy (Mc); Ny Alesund in Norway (Ny); Westford in the USA (Wf); Wettzell in Germany (Wz); Onsala60 in Sweden (On); Badary in Russia (Bd); Ishioka in Japan (Is); Kunming in China (Km); Sheshan in China (Sh); Svetloe in Russia (Sv); Zelenchukskaya in Russia (Zc); Katherine in Australia (Ke). 
Col.~11 -- 1. Astrogeo database; 2. \cite{2012A&A...544A..34P}; 3. This paper.
}
\end{table}

The already calibrated archival VLBI visibility data downloaded from the Astrogeo database for the first four epochs (Table~\ref{tab1}) were also analysed in {\sc Difmap}. Hybrid mapping procedures were performed, including iterations of {\sc clean}ing and phase-only self-calibration. Antenna gain correlation, and self-calibration of phases and amplitudes were then applied to the data. The final images (Fig.~\ref{Fig1}) were produced by applying natural weighting to the visibility data. Based on the self-calibrated visibilities, we used a circular Gaussian brightness distribution model to fit the brightest component in each map. This was considered the `core' of the source, i.e. the optically thick section of the jet at the given frequency. Further along the jet, other features were seen at most of the epochs, located on the western side of the core (Figs.~\ref{Fig1}-\ref{Fig2}). 

To track the motion of the jet components, we used several circular Gaussian brightness distribution model components to fit the visibility data in {\sc Difmap}. Wherever the diameter of the fitted circular Gaussian became smaller than the minimum resolvable size \cite{2005AJ....130.2473K} with the VLBI array, it was replaced by a point model. During the model-fitting process across all available epochs, we followed the strategy of simplicity and consistency from epoch to epoch. The uncertainties of the fitted component positions are estimated in a conservative way, $\sigma_{\rm pos}$ = $\frac{D_{\rm beam}}{2 SNR}$, where $D_{\rm beam}$ is the size of the restoring beam (full width at half-maximum, FWHM), and $SNR$ is the signal-to-noise ratio of the corresponding component. For the uncertainties of the remaining parameters, we give the statistical imaging errors introduced by Fomalont \cite{1999ASPC..180..301F}. Additional 5\% calibration uncertainties were considered for both the image peak intensities and the fitted total flux densities. The parameters obtained from model fitting for J1430+4204 at 8.3/8.6~GHz are listed in Table~\ref{tab2}.

\begin{table}
	\begin{minipage}[]{100mm}
		\caption[]{Fitted model parameters for J1430+4204 at 8.3/8.6~GHz\label{tab2}}\end{minipage}
	\setlength{\tabcolsep}{2.5pt}
	\small
	\centering
	\begin{tabular}{ccccccccc}
		\hline\noalign{\smallskip}
		Epoch & Component & $S_{\rm peak}$ & $S_{\rm int}$ & $R$ & $\phi$ & $\varphi$  & $T_{\rm b}$ & $\delta$\\
		YYYYMMDD&  & (mJy beam$^{-1}$) & (mJy) & (mas) & ($^{\circ}$)& (mas) &($\times$10$^{10}$K) & \\
		(1)& (2) & (3) & (4) & (5) & (6)& (7) &(8)&(9)  \\
		\hline\noalign{\smallskip}
		19960608 & C & 182.3$\pm$9.1  & 174.9$\pm$8.8  & --  & -- & 0.12$\pm$0.01& 131.4$\pm$10.0&26.3$\pm$2.0	\\
		... & J1 & 14.4$\pm$1.0  & 19.3$\pm$1.5  & 0.505$\pm$0.029 & $-$119.3$\pm$ 3.3 & 0.72$\pm$0.06&--&-- \\
		20010409 & C & 65.7$\pm$3.3  & 66.1$\pm$3.3  & --  & --  & 0.07$\pm$0.01 & 120.0$\pm$9.8& 24.0$\pm$2.0\\
		... & J1 & 6.2$\pm$0.4  & 6.1$\pm$0.5  & 0.660$\pm$0.017  & $-$106.9 $\pm$1.5 & 0.16$\pm$0.04$^\star$&--&-- \\
		... & J2 & 3.4$\pm$0.3  & 6.4$\pm$0.7  & 1.657$\pm$0.032  & $-$135.2$\pm$1.1  & 0.88$\pm$0.06&--&-- \\
		20030917 & C & 152.8$\pm$7.6  & 155.2$\pm$7.8  & --  &--  & 0.09$\pm$0.01 & 178.3$\pm$11.6&35.7$\pm$2.3 \\
		... & J1 & 3.0$\pm$0.3  & 2.6$\pm$0.4  & 0.613 $\pm$0.067 & $-$117.3 $\pm$6.2 & 0.40$\pm$0.13$^\star$ &--&--\\
		... & J2 & 2.3$\pm$0.3  & 5.7$\pm$0.9  & 1.842 $\pm$0.088 & $-$127.8$\pm$2.7  & 1.17$\pm$0.18&--&-- \\
		20120108 & C & 182.2$\pm$9.1  & 181.2$\pm$9.1  & --  &--  & 0.18 $\pm$0.01& 53.8$\pm$2.8&10.8$\pm$0.6  \\
		... & J1 & 7.0$\pm$0.4  & 11.2$\pm$0.7  & 0.712 $\pm$0.027 & $-$101.1$\pm$2.2  & 1.09$\pm$0.05 &--&--\\
		... & J2 & 1.1$\pm$0.3  & 2.0$\pm$0.5  & 2.921$\pm$0.167  & $-$104.4 $\pm$3.3 & 1.11$\pm$0.33 &--&-- \\
		20180501 & C & 165.0 $\pm$8.3 & 171.2$\pm$8.6  & --  &--  & 0.15$\pm$0.01 & 71.8$\pm$4.0 &14.4$\pm$0.8 \\
		... & J1 & 3.1$\pm$0.5 	 & 3.2$\pm$0.7  & 0.854$\pm$0.066  & $-$89.1$\pm$4.4  & 0.33$\pm$0.13$^\star$ &--&--\\
		\noalign{\smallskip}\hline	
	\end{tabular}
    \\ {Col.~2 -- component label (C: core, J1 and J2: jet components); Col.~3 -- peak brightness; Col.~4 -- integrated flux density; Col.~5 -- radial separation of the component form the core; Col.~6 -- position angle with respect to the core; Col.~7 -- FWHM size of the fitted Gaussian component; Col.~8 -- core brightness temperature;  Col.~9 -- Doppler factor.}
    \\ {$^\star$ The sizes of point models are estimated using the minimum resolvable sizes \cite{2005AJ....130.2473K}.}
\end{table}

\begin{table}
\begin{center}
\caption[]{The fitted jet component proper motions for J1430+4204 (mas\,yr~$^{-1}$)}\label{tabpm}
 \begin{tabular}{ccccc}
  \hline\noalign{\smallskip}
Comp &  $|\mu_{||}|$      & $|\mu_{\perp}|$ & $|\mu_{\rm tot}|$  & $\mu_{\rm sep}$         \\
      (1)  & (2) & (3) & (4) &(5)\\
  \hline\noalign{\smallskip}
J1  &  0.009$\pm$0.002& 0.015$\pm$0.002   & 0.017$\pm$0.002 & 0.011$\pm$0.002\\ 
J2  &   0.100$\pm$0.017   &   0.120$\pm$0.014  &   0.156$\pm$0.015&  0.112$\pm$0.015 \\
  \noalign{\smallskip}\hline
\end{tabular}
\end{center}
{Col.~2 -- fitted proper motion parallel with the 2.3 GHz jet direction; Col.~3 -- fitted proper motion perpendicular to the 2.3 GHz jet direction; Col.~4 -- proper motion after vectorial addition; Col.~5 -- fitted separation rate assuming simple outward radial motion and not curved trajectory}
\end{table}

\section{Relativistic beaming parameters} \label{app:3}

To evaluate the beaming effect form the innermost region of a relativistic jet, the most common method involving VLBI data is to estimate the core brightness temperature ($T_{\rm b}$). High values of $T_{\rm b}$ indicate Doppler-boosted radio emission in the relativistic jet. Without relativistic beaming, the upper limit of the $T_{\rm b}$ in AGN cores is $\approx 10^{12}$~K caused by inverse Compton scattering \cite{1969ApJ...155L..71K}. A lower intrinsic value of the brightness temperature, $T_{\rm b,int} \approx 5\times10^{10}$~K, was derived under the assumption that there is equipartition between the energy densities of the particles and the magnetic field \cite{1994ApJ...426...51R}. Observed brightness temperatures that exceed the equipartition value are usually attributed to the Doppler-boosting effect. The Doppler factor can then be calculated as
\begin{equation}
\delta_{\rm eq}=\frac{T_{\rm b}}{T_{\rm b,int}}.
\end{equation}
The brightness temperature of the VLBI core component is obtained using the following formula:
\begin{equation}
\rm {T_{\rm b}=1.22 \times 10^{12}(1+z)\frac{S_{\nu}}{\vartheta^2 \nu^{2}} \,\,\, {\rm K}},
\end{equation}
where S$_{\nu}$ is the flux density in Jy at the observing frequency $\nu$, and $\vartheta$ is the FWHM diameter of the circular Gaussian component measured in mas. The observing frequency $\nu$ is in GHz. If we take the values in Table~\ref{tab2} and the equipartition brightness temperature as the intrinsic value, then the calculated Doppler factor $\delta_{\rm eq}$ of the compact core in J1430+4204 varies in the range of 11--36 at the different epochs.

If we use the projected proper motion of J2 (with an absolute value about 0.100~mas\,yr$^{-1}$, Table~\ref{tabpm}) as the apparent bulk motion, the corresponding apparent transverse speed is $\sim 12.5c$. Parameters of the radio jets that characterise their orientation and relativistic motion are the viewing angle with respect to the line of sight, $\theta$, and the Lorentz factor, $\Gamma$. They can be derived from the following formulae (e.g. \cite{1995PASP..107..803U}):
\begin{equation}
 \Gamma=\frac{\beta^2+\delta^2+1}{2\delta}
\end{equation}
\begin{equation}
 \theta=\arctan{\frac{2\beta}{\beta^2+\delta^2-1}},
\end{equation}
where $\beta=12.5$ is the apparent transverse speed measured in the units of $c$. We apply the average of the $\delta_{\rm eq}$ values derived for J2 at the different epochs as the Doppler factor, $\delta \approx 22$. This way we obtain $\Gamma \approx 14.6$ for the Lorentz factor and $\theta \approx 2.2^{\circ}$ for the jet viewing angle.

\end{document}